\documentclass[useAMS,usenatbib,onecolumn,usegraphicx]{mn2e}

\usepackage{epsfig}
\usepackage{amssymb}
 
\def\rund#1{\left( #1 \right)}

\def\dvk{\frac{{\rm d}^3k}{(2\pi)^3}}

\def\dvx{{\rm d}^3x}
\def\dtx{{\rm d}^2x_{\perp}}
\def\dpx{{\rm d}x_{\parallel}}
\def\x{\bmath{x}}
\def\k{\bmath{k}}
\def\l{\bmath{l}}
\def\vtheta{\bmath{\theta}}
\def\xt{\bmath{x}_{\perp}}
\def\kt{\bmath{k}_{\perp}}
\def\xp{{x}_{\parallel}}
\def\kp{{k}_{\parallel}}

\def\Mpc{\rm Mpc}

\def\mK{\rm mK}

\def\gradt{{\bmath{\nabla}_{\perp}}}

\newcommand{\beq}{\begin{equation}}
\newcommand{\eeq}{\end{equation}}
\newcommand{\bal}{\begin{aligned}}
\newcommand{\eal}{\end{aligned}}
\newcommand{\beqa}{\begin{eqnarray}}
\newcommand{\eeqa}{\end{eqnarray}}

\title{Precision of diffuse 21-cm lensing
}

\author[Lu, \& Pen]{
Tingting Lu,$^{1}$\thanks{E-mail:\ ttlu@cita.utoronto.ca}
Ue-Li Pen,$^{2}$\thanks{E-mail:\ pen@cita.utoronto.ca}
\\
$^1$\,Department of Astronomy and Astrophysics, University of Toronto, M5S
3H4, Canada \\
$^2$\,Canadian Institute for Theoretical Astrophysics, University of
Toronto, M5S 3H8, Canada \\
}

\begin{document}

\date{version 1 October 2007}

\pagerange{\pageref{firstpage}--\pageref{lastpage}}
\pubyear{2007}

\maketitle

\label{firstpage}

\begin{abstract}	

We study the limits of accuracy for weak lensing maps of dark matter
using diffuse 21-cm radiation from the pre-reionization epoch
using simulations.  We improve on previous ``optimal'' quadratic
lensing estimators by using shear and convergence instead of
deflection angles.  We find that non-Gaussianity provides a limit to
the accuracy of weak lensing reconstruction, even if instrumental
noise is reduced to zero.  The best reconstruction result is equivalent to
Gaussian sources with effective independent cell of side length
$2.0h^{-1}\, \rm Mpc$.  
 Using a source full map from z=10-20, this
limiting sensitivity allows mapping of dark matter at a Signal-to-Noise 
ratio (S/N) greater than 1 out to
$l\lesssim 6000$, which is better than any other proposed technique
for large area weak lensing mapping.

\end{abstract}
 

\section{Introduction} \label{INTRO}

The lens mapping of dark matter is an essential cornerstone of
modern precision cosmology.  Weak gravitational lensing has developed
rapidly over the past years, which allows the measurement of the
projected dark matter density along arbitrary lines-of-sight
using galaxies as sources.  Recently, \citet{2007PhRvD..76d3510S} have
demonstrated the first CMB lensing detection.  The goal is now to
achieve high precision cosmological measurements through lensing, at
better than 1\% accuracy.

Galaxies are plentiful on the sky, but their intrinsic properties are
not understood from first principles, and must be measured from the
data.  Future surveys may map as many as $10^{10}$ source objects.
Using galaxies as lensing sources has several potential limits
\citep{2004PhRvD..70f3526H}, including the need to calibrate redshift
space distributions and PSF corrections, to be better than the desired
accuracy, say 1\%.  This will be challenging for the next generation
of experiments.

Some sources, such as the CMB, are in principle very clean, since its
redshift and statistical properties are well understood.
Unfortunately, there is only one 2-D CMB sky with an exponential
damping at $l \gg 1000$, which limits the number of source modes to
$\sim 10^6$.

The potential of detecting the 21-cm background from the dark ages
will open a new window for cosmological detections. Studying the 21-cm
background as high redshifts lensing source, as well as the physics of
the 21-cm background itself, provide rich and valuable information to
the evolution of universe.  The number of modes on the sky is
potentially very large, with numbers of $10^{16}$ or more.  For this
reason, 21-cm lensing has recently attracted attention. However, most
of the reconstruction methods are based on a Gaussian assumption
\citep{2004NewA....9..417P,2004NewA....9..173C,2006ApJ...653..922Z,
  2006astro.ph.11862B,2007arXiv0706.0849H}. In contrast to CMB
lensing, where the Gaussian assumption works well, non-Gaussianity in
21-cm lensing may affect the results.  Non-linear gravitational
clustering leads to non-Gaussianity, and ultimately to reionization.
In this paper, we will address the problem of the lensing of
pre-reionization gas.

21-cm emission is similar to CMB: both are diffuse backgrounds.  It is
natural to apply the techniques used in CMB lensing.
\citet{2002ApJ...574..566H} expand the CMB lensing field in terms of
the gravitational potential (or deflection angles), and construct a
trispectrum based quadratic estimator of potential with maximum S/N.
However, unlike CMB, the 21-cm background has a 3-D distribution and is
intrinsically non-Gaussian.  A fully 3-D analysis is explored in
\citet{2006ApJ...653..922Z}, where they generalize the 2-D quadratic
estimator of CMB lensing \citep{2002ApJ...574..566H} to the 3-D
Optimal Quadratic Deflection Estimator (OQDE).  

A local estimator was proposed
in \citet{2004NewA....9..417P}, which assumed a power law density
power spectrum. In this paper, we will design localized estimators for
the lensing fields under the Gaussian assumption, and apply the
derived reconstruction technique to Gaussian and non-Gaussian
sources. The influence of non-Gaussianity can be measured by comparing
the numerical results between the Gaussian sources and non-Gaussian
sources.

Quadratic lensing reconstruction is a two point function of the lensed
brightness temperature field of the 21-cm emission.  In the paper, 3-D
quadratic estimators are constructed for the convergence ($\kappa$),
as well as the shear ($\gamma$).  Our method recovers the $\kappa$ and
$\gamma$ directly instead of gravitational potential or deflection
angles. Our estimators have in principle the same form as the OQDE,
consisting of the covariance of two filtered temperature maps. The OQDE  
reconstructs the
deflection angle, while our estimators reconstruct the kappa and
shear fields. Our filtering process can be written as a convolution of
the observed fields. As presented in Appendix and section 4, our
combined estimator is unbiased, and equally optimal as the OQDE for Gaussian
sources, and has better performance for non-Gaussian sources, and
recovers three extra (constant) modes.

Other authors also developed reconstruction methods from alternative
approaches.  \citet{2006astro.ph.11862B} give a estimator for
shear. They choose the separate 2-D slices at certain redshift
intervals, and then these slices can be treated as independent samples
for the same lensing structure. As a result, the information between
these slices are lost. \citet{2004NewA....9..173C} expands the
lensed field to a higher order of the gravitational potential, and
investigates the higher order correction to the lensed power
spectrum. 

The paper is organized as follows: The basic framework of lensing and
the reconstruction method is introduced in $\S 2$. The numerical
methods are presented in $\S 3$. The results are discussed in $\S
4$. We conclude in $\S 5$.

\section{Lensing and reconstruction}

Photons are deflected by clumpy matter when they propagate from the
source to the observer. This effect can be used to map the mass
distribution if we can measure the distortion of an image. In this
section, we will first review the lensing theory, which serves to
define our notation.  We then develop an optimal quadratic estimator
using a maximum likelihood method.  The reconstruction depends on the
power spectrum of the source. The noise and normalization of the
reconstruction are calculated in the appendix.

\subsection{Lensing} \label{LENS}

The Jacobian matrix describing the mapping between the source 
and image planes is defined as 
\beq
{\mathbf J} (\vtheta,\chi)
={1\over f_K(\chi)}{{\rm \partial} {\bmath x} \over {\rm \partial} \vtheta} \ .
\eeq
Here $\chi$ is the radial coordinate, and $f_K(\chi)$ is the comoving angular
diameter distance. We consider a ray bundle intersecting at the observer and 
denote $\bmath x(\vtheta,\chi)$ as the comoving transverse coordinate of a ray.

In the lensing literature, the physical quantities frequently used to
describe a lensing field are convergence $\kappa$ and shear $\gamma$,
which are given by
\begin{displaymath}
{\mathbf J}(\vtheta,\chi)=
\left(\begin{array}{cc}
1-\kappa-\gamma_1 & -\gamma_2 \\
-\gamma_2         & 1-\kappa+\gamma_1
\end{array}\right) \ .
\end{displaymath}
Equivalently, the convergence and shear can also be written as $\kappa=(\Phi_{,11}+
\Phi_{,22})/2\ ; \ \gamma_1=(\Phi_{,11}-\Phi_{,22})/2\ ; \ \gamma_2=\Phi_{,12}$. $\Phi$ 
is the projected 2-D potential:
\beq
\Phi 
= {2\over c^2 }\int_0^{\chi} {\rm d}\chi' {f_K(\chi')f_K(\chi-\chi') \over f_K(\chi)} 
\phi[f_K(\chi')\bmath \theta(\chi'),\chi'] \ .
\eeq
Here subscripts '1' and '2' refer to the derivative to the two perpendicular transverse coordinates, and $\phi$ is the 3-D Newtonian gravitational potential.  
Note that the integral is along the actual perturbed path of each photon. 
In the Born approximation, the deflection is approximated by an
integral along the  
unperturbed path.
 
In the small angle approximation \citep{1954ApJ...119..655L}, $
\nabla^2_{\perp}$ can  
be replaced by $\nabla^2 $ in the integral.  
We  get the Limber equation
\beq
\kappa={ 3  H_0^2 \over 2 }\ \Omega_{\rm m}\ \int_0^{\chi} {\rm d}\chi' g(\chi',\chi)\ {\delta \over a(\chi')} \ ,
\label{eq:Kproj}
\eeq
with $g(\chi',\chi)={f_K(\chi')f_K(\chi-\chi') / f_K(\chi)}$. $\Omega_{\rm m}$ is 
the mass density parameter, $H_0$ is the current Hubble constant, $a$ is the scale factor, 
and $\delta$ is the over-density.

\citet{1992ApJ...388..272K} derived the Fourier-space version of the Limber equation
\beq
P_{\kappa}(l) =
{9\over 4}\rund{H_0\over c}^4\Omega_{\rm m}^2
\int_0^{\chi_{\rm H}}{\rm d}\chi{g^2(\chi)\over a^2(\chi)}
P\rund{{l\over f_K(\chi)},\chi} \ .
\label{eq:Limber}
\eeq
Here $P_{\kappa}(l)$ is the 2-D power spectrum of the $\kappa$ field,
$P(l/f_K(\chi),\chi)$ is the 3-D power spectrum of matter, 
and $\chi_{\rm H}$ is the comoving distance 
to the  Hubble horizon. 
The equation is valid when the power spectrum $P_{\kappa}$ evolves slowly over
time corresponding to the scales of fluctuation of interest, and these 
fluctuation scales are smaller than the horizon scale. 

\subsection{Reconstruction of large-scale structure} \label{RC}

We first heuristically review the quadratic lensing estimation in two
dimensions.  Then we will proceed with a generalization to 3-D with a
quantitative derivation.

Lensing will change the distribution of a temperature field by changing
length scales.  Lensing estimation relies on statistical changes to
quadratic quantities in the source plane temperature field.
We use a tilde to denote a lensed  quantity.
All estimators work by convolving the temperature field with a window, 
\beq
\tilde T_1(\x)=\int {\rm d}^2x' \tilde T(\x') W_1(\x-\x') \ ,
\eeq
and a second window
\beq
\tilde T_2(\x)=\int {\rm d}^2x' \tilde T(\x') W_2(\x-\x') \ .
\eeq
The quadratic estimator is simply the product of the two convolved
temperature fields,
\beq
E(\x) \equiv \tilde T_1(\x) \tilde T_2(\x).
\eeq
In the weak lensing case, the estimator is a linear function of
the weak lensing parameters ($\kappa,\gamma$).
The simplest case is two equal, azimuthally symmetric window functions
$W_1=W_2=f(r)$.  
Considering the limit that $\kappa$ is a constant value,
the estimator is linearly proportionate 
to $\kappa$: 
\beq
\langle E \rangle \propto  \kappa + V \ ,
\label{eqn:e}
\eeq
and $V$ is the mean covariance.
Here $\langle...\rangle$ means ensemble average. For a stochastic random field, the
ensemble average can be calculated by the volume average if the volume is big enough.
We can absorb $V$ as well as the normalization coefficient into $E$ for convenience, 
i.e., $E(\x) \equiv \tilde T_1(\x) \tilde T_2(\x)-V$. 
When $\kappa$ is spatially variable, $E$ needs to be
normalized by a scale dependent factor $b(k)$.  This corresponds to a convolution of $\kappa$
with a kernel:  
\beq
\langle{E(\x)}\rangle = \int {\rm d}^2x' \kappa(\x') b(\x-\x') \ ,
\label{eqn:bias}
\eeq
where kernel $b$ is the Fourier transform of the normalization factor.

One can optimize the functions to minimize the error on the lensing
variables.  In this paper we will compare various forms of the
smoothing windows, which include as special case the traditional
Optimal Quadratic Deflection Estimator.
The simplest case is a constant value of $\kappa$, for which one can
compute its variance
\beq
\langle \kappa^2 \rangle = \langle{(\tilde T_1(\x)^2)^2}\rangle .
\label{eq:4pt}
\eeq
Lensing is a small perturbation of the variance, therefore we can
calculate the variance  
from the unlensed source field, i.e., $\langle{(T_1(\x)^2)^2}\rangle \approx
\langle{(\tilde T_1(\x)^2)^2}\rangle  $.
Performing a variation to minimize the variance, one can find the
optimal window function.
It turns out that the window functions do not depend on the spatial
structure of the lensing field.  Only the normalization factor $b$ in
Eq. (\ref{eqn:bias}) is scale dependent.  We solve the optimal window 
function at scales where the constant $\kappa$ approximation works
well, and the solution  should also be optimal for other scales. 

Shear and deflection angles are tensorial and vectorial quantities and
require anisotropic or vectorial choices of the window function 
\beq
{\mathbf E_{\rm \gamma}}={\tilde {\bmath T_1}} {\tilde {\bmath T_2}} ,\,
{\tilde {\bmath{T}_1}}=\int{\rm d}^2{\theta'}{\tilde T(\vtheta')}\bmath{W_1}
(\vtheta-\vtheta') ,\, 
{\tilde {\bmath{T}_2}}=\int{\rm d}^2{\theta'}{\tilde T(\vtheta')}\bmath{W_2}
(\vtheta-\vtheta') \ , 
\eeq
\beq
{\bmath E_{\rm d}}={\tilde {\bmath T_1}} {\tilde T_2} ,\,
{\tilde {\bmath{T}_1}}=\int{\rm d}^2{\theta'}{\tilde T(\vtheta')}\bmath{W_1}
(\vtheta-\vtheta') ,\,  
{\tilde T_2}=\int{\rm d}^2{\theta'}{\tilde T(\vtheta')}W_2(\vtheta-\vtheta') \ .
\label{eq:e_t}
\label{eq:e_vec}
\eeq
This will be explained in detail in sections \ref{EST_S} and \ref{EST_COMB}.

The source is usually treated as a Gaussian stochastic field in the
literature on reconstruction methods. While this is valid for CMB on
large angular scales, 21-cm background sources are not always
Gaussian. In this paper we attempt to understand the influence of this
non-Gaussianity. Optimal estimators for Gaussian sources are not
necessarily  optimal for non-Gaussian sources. Here, we will construct
the convergence and shear field 
directly, instead of following the deflection angles or potential
field reconstruction in CMB lensing.  There are three reasons to do
this: Firstly, the strength of lensing is evident through the magnitude
of $\kappa$ or $\gamma$ since they are dimensionless quantities. The
rms deflection angle of photons from 21-cm emission is at the
magnitude of a few arcmin, which is comparable to the lensing
scales we are resolving. Some authors argued that perturbation theory
on the deflection angle will break down at these scales
\citep{2004NewA....9..173C, 2006ApJ...647..719M}. However, $\kappa$
and $\gamma$ are still small  and can still work with
perturbation calculations without ambiguity.  
Secondly, $\kappa$ and $\gamma$ have well defined limits as they
approach a constant, while only spatially variable deflection angles
or potentials can be measured. This significantly simplifies the
derivations. Finally, $\kappa$ and $\gamma$ are standard
variables to use in broader lensing studies, such as strong lensing
and cosmic shear. Using the same convention in different subfields will help
to generalize the underlying physics of lensing.

The estimators are unbiased, as shown in the appendix.  Furthermore,
we confirm that our combined estimators from $\kappa$ and $\gamma$ have the
same optimality as the OQDE for Gaussian sources. 
When the sources are non-Gaussian, our estimators have better S/N. 

\subsubsection{Maximum likelihood estimator of $\kappa$} \label{EST_K}

We now derive the quantitative window functions for 21-cm lensing
reconstruction.  Due to their similarity, it is helpful to quickly
review the reconstruction in CMB lensing:
The early work by \citet{1999PhRvD..59l3507Z} used the quadratic
combination of the derivatives of the CMB field to reconstruct the
lens distribution.  Since the CMB has an intrinsic Gaussian distribution,
The optimal quadratic estimator
\citep{2001ApJ...557L..79H} can also be applied to lensing
reconstruction with CMB polarization \citep{2002ApJ...574..566H}.
\citet{2006ApJ...653..922Z} generalized the optimal quadratic
estimator of CMB lensing to 21-cm lensing.  

We will construct estimators for $\kappa$ and $\gamma$ with the 21-cm
brightness temperature fields, starting from a maximum likelihood
method which is consistent with the quadratic minimum variance method
when the field is Gaussian.  We will show that the OQDE
and our approach are the same if the sources are Gaussian, however the
problem is simplified in a intuitive way by using the limit that
$\kappa$ and $\gamma$ vary slowly in small scales.

The magnification is
\beq
\mu=\frac{1}{(1-\kappa)^2-\gamma^2} \sim 1+2 \kappa \ .
\eeq
The last approximation is valid 
since both $\kappa$ and $\gamma$ are much smaller than $1$ in the weak
lensing regime. 

We use Bayesian statistics and assume the prior distribution of parameter
$\kappa$ to be flat. For a $M$ pixel map on the sky, the posterior
likelihood function of the source field has a Gaussian distribution,
and can be written as
\beq
{\cal P}(\tilde T(\k)) = (2\pi)^{-M/2} {\det
(\mathbf C_{\tilde T \tilde T})}^{-{1\over 2}} e^{ -{1\over 2} \tilde T^{\dagger}
{\mathbf C_{\tilde T \tilde T}}^{-1}\tilde T} \ .
\label{eq:pr2}
\eeq
Here $\tilde T=\tilde T_b+n$ is the brightness temperature of the diffusive 21-cm 
emission lensed by the large-scale structure plus measurement noise. 
To simplify the algebra, we  use the negative logarithm ${\cal L}$ of the 
likelihood function in our calculation,
\beq
{\cal L} = -\ln {\cal P}={1\over 2} \tilde T^{\dagger} \mathbf C_{\tilde T
\tilde T}^{-1} \tilde T + {1\over 2}\ln\det
\mathbf C_{\tilde T \tilde T}.
\label{eq:lform}
\eeq
Here $\tilde T$ is the 3-D discrete Fourier transform of measured temperature.
$\mathbf C_{\tilde T \tilde T} = \mathbf C_{\rm S} + \mathbf C_{\rm N}$ is the 
covariance matrix, and the signal contribution $\mathbf C_{\rm S} $ and noise contribution
$ \mathbf C_{\rm N} $ are both diagonal in Fourier space and uncorrelated to each other.
In the continuum limit, the likelihood function can be written as 

\beq
{\cal L}
={1\over 4 \pi^2} \lbrack\int {\rm d}^3 k \ln{\tilde P^{\rm tot}_{\rm 3D}({\k})} 
+ \int {\rm d}^3 k \frac{|\tilde T(\k)|^2} {\tilde P^{\rm tot}_{\rm 3D}({\k})}\rbrack \ .
\eeq
We use $\tilde P^{\rm tot}_{\rm 3D}=\tilde P_{\rm 3D}({\k})+P_{\rm N}(\k)$ to represent 
signal plus noise power spectrum in the following text, where
$\tilde P_{\rm 3D}({\k})$ is 3-D power spectrum of the
distorted 21-cm field, and $P_{\rm N}(\k)$ is the noise power spectrum. 
The dimensionless power spectrum of the 3-D 21-cm gas slices can be written as 
\beq
\Delta^2_{\rm 3D}(k)= { k^3 \over 2\pi^2} P_{\rm 3D}(k) \ ,
\eeq
where $k=|\k|$ since the gas is statistically isotropic.

The geometry of the 21-cm field will be changed by the lensing:    
\beq
\tilde T_b(\kt, \kp ) 
=  \int {\rm d}^3x \tilde T_b(\bmath x) e^{-{\rm i} \k \cdot \bmath x}  
=  \int {\rm d}^2x_{\perp}\int {\rm d} x_{\parallel} T_b((1-\kappa)\bmath x_{\perp},x_{\parallel}) 
   e^{-{\rm i} (\kt \cdot \bmath x_{\perp}+\kp x_{\parallel} )}  
=  {1\over(1-\kappa)^2}T_b((1+\kappa)\kt,\kp) \ ,
\eeq
where '$\perp$' and '$\parallel$' mean the perpendicular and parallel
direction of  the line-of-sight respectively. We ignore the
contribution of shear first.   
Then the length scale is magnified on the transverse plane by a
factor  
$\kappa$. Isotropy is broken in 3-D  but  is still
conserved on the 2-D cross section.
The statistical properties of the 21-cm field will be changed by the
lensing, i.e.,  the power spectrum will also change:    
\beq
\langle{\tilde T_b^{\ast}(\kt,\kp)\tilde T_b(\kt',\kp')}\rangle
= (2\pi)^2\delta^{\rm 2D}(\kt-\kt') (2\pi)\delta^{\rm D}(\kp-\kp')\tilde P_{\rm 3D}(\kt,\kp) \ .
\label{eq:lens_ulens}
\eeq
The delta function has the property:
\begin{equation}
\delta^{\rm 2D}((1+\kappa)\kt-(1+\kappa)\kt')={1\over(1+\kappa)^2}\delta^{\rm 2D}(\kt-\kt') \ .
\end{equation}
Therefore the relationship between the unlensed and lensed power spectrum is 
\beq
\tilde P_{\rm 3D}(\kt,\kp)
= (1+2\kappa)P_{\rm 3D}((1+\kappa)\kt,\kp) 
= (1+2\kappa)P_{\rm 3D}(\sqrt{(1+\kappa)^2k_{\perp}^2+\kp^2}) 
\approx(1+2\kappa)(P_{\rm 3D}(k)+\kappa \Delta P_{\rm 3D}) \,,
\eeq
where $\Delta P_{\rm 3D}=P_{\rm 3D}'k({k_{\perp}^2/k^2})$, and $P_{\rm 3D}'(k)=
{\rm d} P_{\rm 3D}(k)/ {\rm d} k$. The second equivalence 
is due to the statistical isotropy of the unlensed power spectrum.

Differentiation of the lensed power spectrum gives
\begin{equation}
\frac{{\rm \delta}{\tilde P_{\rm 3D}(\k)}}{{\rm \delta} \kappa}=2P_{\rm 3D}+(1+2\kappa)\Delta P_{\rm 3D} \ ,
\end{equation}
and the maximum likelihood condition requires 
\beq
\frac{{\rm \delta}{\cal L}}{{\rm \delta} \kappa}\approx{1\over2}L^3\int {\dvk}
\frac{(\tilde P^{\rm tot}_{\rm 3D}-|\tilde T|^2L^{-3}) }
{{P^{\rm tot}_{\rm 3D}}^2} {{\rm \delta}\tilde P_{\rm 3D}\over {\rm \delta} \kappa}  =0 \,,
\label{eq:maxl}
\eeq
which has the solution
\beq
E_{\rm \kappa}=
\int{\dvk} (|\tilde T|^2L^{-3}){\cal F}^{\rm \kappa}(\k)-V_{\rm \kappa}
 \ .
\label{eq:est_kappa}
\eeq
We have approximated $\tilde P^{\rm tot}_{\rm 3D}$ by $P^{\rm tot}_{\rm 3D}$ in 
the denominator of Eq. (\ref{eq:maxl}). To simplify the problem, we assume the source is
a cube with physical length $L$ in each dimension. 
The offset constant $V_{\rm \kappa}=\langle{\sigma^2}\rangle=\int {{\rm d}^3 k/ (2\pi)^3} P^{\rm tot}_{\rm 3D}(\k)
{\cal F}^{\rm \kappa}(\k)$,  
and the optimal filter ${\cal F}^{\kappa}$ is  
\beq
{\cal F}^{\rm \kappa}(\k)=\frac{2P_{\rm 3D}(\k)+ \Delta P_{\rm 3D}(\k)}{{P^{\rm tot}_{\rm 3D}}^2(\k)Q_{\rm \kappa}} \ ,
\label{eq:wk_pk}
\eeq
with $Q_{\rm \kappa}=\int {{\rm d}^3 k/(2\pi)^3} (2P_{\rm 3D}+\Delta P_{\rm 3D})(\k)
{\cal F}^{\rm \kappa}(\k)$. 

From  Parseval's theorem, we can rewrite Eq. (\ref{eq:est_kappa}) in the 
form of a convolution of the density field and a window function in real space
\beq
\int {\dvk}\tilde T^{\ast}(\k) \tilde T(\k){\cal F}^{\rm \kappa}(\k)
= \int {\dvx}\tilde T^{\rm \kappa}_{\rm w_1}(\x) \tilde T^{\rm \kappa}_{\rm w_2}(\x) 
=  L^2 \int \dpx \tilde T^{\rm \kappa}_{\rm w_1}(\xt,x_{\parallel}) 
   \tilde T^{\rm \kappa}_{\rm w_2}(\xt,x_{\parallel}) \,.
\label{eq:parseval}
\eeq
In Eq. (\ref{eq:parseval}) the two window functions are the
decomposition of the optimal filter
$W^{\rm \kappa}_1(\k)W^{\rm \kappa}_2(\k)={\cal F}^{\rm \kappa}(\k)$.
The last '$=$' in Eq. (\ref{eq:parseval}) holds when $\kappa$ is
constant. One can choose $W^{\rm \kappa}_1(\k)=W^{\rm \kappa}_2(\k)=\sqrt{{\cal F}_{\rm \kappa}}$. If ${\cal F}_{\rm \kappa}<0$, we choose $W^{\rm \kappa}_1=-W^{\rm \kappa}_2=\sqrt{|{\cal F}_{\rm \kappa}}|$.
The convergence field is equivalent to the covariance of the measured maps 
with two windows applied.  In the slowly spatially varying $\kappa$ limit, all
decomposition into two windows are equivalent.   
As we will show later, the shear construction can also be represented in the form of 
the covariance of two filtered temperature maps. These maps
will have symmetric Probability Density Function
 (PDF), which can reduce the non-Gaussianity of the maps so that a better S/N 
level can be achieved, when the shear window functions are chosen properly.
The last two steps in Eq. (\ref{eq:parseval}) assumes the fluctuation of the 
convergence field is slow compared to the filter.  Then
we can apply the estimator to each beam in the map: 
\begin{equation} 
E_{\rm \kappa}(\xt)
= L^{-1} \int{\rm d} \xp \tilde T^{\rm \kappa}_{\rm w_1}(\x) \tilde T^{\rm \kappa}_{\rm w_2}(\x)-V_{\rm \kappa} \,,
\label{eq:kappax}
\end{equation} 
where $\tilde T^{\rm \kappa}_{\rm w_1}$ and $\tilde T^{\rm \kappa}_{\rm w_2}$ are the
convolution of $\tilde T$  
and window function $W^{\rm \kappa}_1(\x)$ and $W^{\rm \kappa}_2(\x)$
respectively, which are the real  
space version of $W^{\rm \kappa}_1(\k)$ and $W^{\rm \kappa}_2(\k)$. 
The reconstruction of the $\kappa$ is dominated by the gradient of the 
power spectrum ${\rm d}\ln{\Delta^2}/{\rm d}\ln{k}$, which follows the
expression of our  estimator in Eq. (\ref{eq:est_kappa}).

We can then generalize the estimator to a spatially varying lensing field. 
In the appendix we show  
\beq
\int {\rm d}^2 x_{\perp}' \kappa(\xt') b_{\rm \kappa}(\xt-\xt') 
= \langle{ E_{\rm \kappa}({\xt})}\rangle    \ .  
\eeq
Equivalently, for smaller scales, we will need to normalize the reconstructed lensing 
field by a scale
dependent factor in Fourier space, which is calculated in the appendix. 
\beq
\hat \kappa({\bmath l})=b_{\rm \kappa}^{-1}({\bmath l}) E_{\rm \kappa}({\bmath l})
= \kappa(\bmath l)+n(\bmath l) \ ,
\label{eq:bias}
\eeq 
where ${\bmath l}=\kt \chi(z_{\rm s})$, and $z_{\rm s}$ is the redshift of the source.
Here $b_{\rm \kappa}(l)$ is the normalization factor ($\lim_{l\to 0}b_{\rm \kappa}(l)=1$), 
and $n(l)$ is the
noise, since different Fourier modes are independent.
They do not depend on direction because variables related to $\kappa$ are 
isotropic on the transverse plane.  
In the appendix,  we show that the normalization factor 
is unity at small $l$ when $Q_{\rm \kappa}$ has the form as
$Q_{\rm \kappa}=\int {{\rm d}^3 k/(2\pi)^3} (2P_{\rm 3D}+ 
\Delta P_{\rm 3D})(\k) {\cal F}^{\rm \kappa}(\k)$. 

\subsubsection{Estimator of shear}   \label{EST_S}

When  shear is taken into account, not only the scale but the directions
of the coordinates are changed. We will start the derivation from the
constant shear case.
\beqa
\tilde T_b(\kt,\kp) 
&=&\int {\rm d}^3 x \tilde T_b(\bmath x) e^{-{\rm i} \k \cdot \bmath x} 
=\int {\rm d}^2 {x_{\perp}}\int {\rm d} x_{\parallel} T_b(\mathbf J\bmath x_{\perp},x_{\parallel})  
e^{-{\rm i} (\kt\cdot\bmath x_{\perp}+\kp x_{\parallel} )} \nonumber \\
&=& |\mathbf J|^{-1} \int {\rm d}^2 {x_{\perp}'}\int {\rm d} x_{\parallel} T_b(\bmath x_{\perp}',x_{\parallel})  
 e^{-{\rm i} (\kt'\cdot\bmath x_{\perp}'+\kp x_{\parallel} )} 
= |\mathbf J|^{-1}T_b(\mathbf J^{-1}\kt,\kp) \,,
\eeqa
here ${\rm d}^2 x_{\perp}'=|\mathbf J|{\rm d}^2 x_{\perp},\kt'=\mathbf J^{-1}\kt$.
Now the symmetry is broken even on the transverse plane due to the anisotropic 
distortion caused by the shear.

Since  $\delta^{\rm 2D}(\mathbf J^{-1}\k)=|\mathbf J| \delta^{\rm 2D}(\k)$,
Eq. (\ref{eq:lens_ulens}) implies
\beq
\tilde P_{\rm 3D}(\kt,\kp)
= |\mathbf J|^{-1} P_{\rm 3D}(\mathbf J^{-1}\kt,\kp) 
 \approx (1+2\kappa) [P_{\rm 3D}(k)+ 
 \Delta P_{\rm 3D}(\k)(\kappa+\gamma_1 \cos2\theta_{\kt}+\gamma_2 \sin2\theta_{\kt}) ] \,,
\eeq
where $\theta_{\kt}$ is the angle between $\bmath k_{\perp}$ and the transverse coordinate.

Maximum likelihood requires ${\rm \delta}{\cal L}/{{\rm \delta} \gamma_1}=0$, and ${\rm \delta}{\cal L}/{{\rm \delta} \gamma_2}=0$. 
The maximum likelihood shear estimators can be written as a tensor $\mathbf E_{\rm \gamma}$:
\beq
E_{\rm \gamma i j} = L^{-1} \int{\rm d} \xp  \tilde T^{\rm \gamma}_{\rm w_i} \tilde T^{\rm \gamma}_{\rm w_j} \ ,
\label {eq:tensorx}
\eeq
where $\tilde T^{\rm \gamma}_{\rm w_i}$ is convolution of the temperature field with $W^{\rm \gamma}_{\rm i}$, and $W^{\rm \gamma}_{\rm i}(\k)=(2 \Delta P_{\rm 3D}/P_{\rm 3D}^2 Q_{\rm \gamma})^{1/2}\hat k_{\rm i}$, $\hat k_{\rm i}$ ($i,j=1,2$) is one of the two unit vectors on the transverse plane.  
When $\Delta P<0$, we can choose $W^{\rm \gamma}_{\rm 1}=|2 \Delta P_{\rm 3D}/P_{\rm 3D}^2 Q_{\rm \gamma}|^{1/2}\hat k_{\rm 1},W^{\rm \gamma}_{\rm 2}=-|2 \Delta P_{\rm 3D}/P_{\rm 3D}^2 Q_{\rm \gamma}|^{1/2}\hat k_{\rm 2}$.
The normalization factor $Q_{\rm \gamma}=\int {{\rm d}^3 k/(2\pi)^3} {\Delta P_{\rm 3D}(\k)}
 \hat k_1 \hat k_2 W^{\rm \gamma}_1(\k) W^{\rm \gamma}_2(\k) $.
The two components of shear are now:
\beq
\hat\gamma_1=E_{\rm \gamma 1 2}=E_{\rm \gamma 2 1} , \
\hat\gamma_2={E_{\rm \gamma 1 1}-E_{\rm \gamma 2 2} \over 2} \ . 
\label{eq:shearx}
\eeq

Note that there is a difference between the reconstruction  for
convergence and  
shear. Shear reconstruction depends on the gradient of $P(k)$, 
while convergence reconstruction depends on the gradient of
$\Delta^2(k)$ in a 2-D analogue.  
To test our method, we can generated a Gaussian source field with power 
law power spectrum $P(k)=k^{\beta}$. 
In the 2-D analogue case, the convergence field can not be measured if
$\beta=-2$,   
because the variance is conserved. However in 3-D, when $\beta=-3$, the
convergence field can  
still be measured,  which is due to the more complicated shape of the
window function in 3-D.  When $\beta=0$, the shear can not be measured in
either 2-D or 3-D.

In analogy to $\kappa$ reconstruction, we can calculate the
normalization factors $b_{\rm \gamma_1}$ and $b_{\rm \gamma_2}$.  The
calculations for the normalization factors and noise are presented in
the appendix.

\subsection{The combined estimator and the OQDE}   \label{EST_COMB}

The combined estimator of $\kappa$ can be written as
\beq
\hat \kappa_{\rm comb}(\l)
= \frac{{\hat\kappa(\l)N_{\rm \kappa}(\l)^{-1}}+{\hat \gamma_{\rm E}(\l) N_{\rm \gamma_E}(\l)^{-1}}}
  {N_{\rm \kappa}(\l)^{-1}+N_{\rm \gamma_E}(\l)^{-1}} \ ,
\label{eq:comb}
\eeq
where $\hat \gamma_{\rm E}$ is the convergence constructed from shear field,
\beq
\hat \gamma_{\rm E}(\l)
=\hat \gamma_1(\l) \cos{2\vtheta_l}+\hat \gamma_2(\l) \sin{2\vtheta_l}  \ ,
\eeq
and $\vtheta_l$ is the angle of $\l$. 

The 2-D OQDE in CMB lensing can be written as product of two filtered
temperature field  
\citep{2001ApJ...557L..79H,2006PhR...429....1L}.  
Furthermore, the 3-D OQDE can be written in the same form of
Eq. (\ref{eq:e_vec}), 
though it is  not explicit (private communication with  Oliver Zahn). 
\beqa
{\bmath E_{\rm d}} (\vtheta) = L^{-1} \int{\rm d}\xp \bmath{T}_1(\vtheta,\xp) T_2(\vtheta,\xp) \ ,
\eeqa
and
\beqa
\int {\rm d}^2 \theta' {\bmath d}(\vtheta')b_{\rm d}(\vtheta-\vtheta')= \langle{\bmath E_{\rm d}(\vtheta)}\rangle \ .
\eeqa
$b_{\rm d}$ is a normalization factor, $\bmath{T}_1=\int{\rm d}^2{\theta'}T(\vtheta')\bmath{W_1}(\vtheta-\vtheta')$ 
and $T_2=\int{\rm d}^2{\theta'}T(\vtheta')W_2(\vtheta-\vtheta')$ are
convolved temperature fields, where the window functions are Fourier
transforms of: 
\beqa
\bmath{W}_1(\l, \kp) 
&=& {- {\rm i}\l P_{\rm 3D}(\l,\kp) \over \tilde P^{\rm tot}_{\rm 3D}(\l,\kp)}, \nonumber\\
W_2(\l,\kp) &=& {1 \over \tilde P^{\rm tot}_{\rm 3D}(\l,\kp)} \ .
\eeqa
We note that the OQDE and our estimators have the same form.
The contribution from lensing in Eq. (\ref{eq:4pt}) is secondary,
and the noise of reconstruction is mainly determined by the unlensed
terms. Therefore we can measure the numerical reconstruction noise
without  lensing  the sources.

\section{Numerical methods}    \label{NUM}

\subsection{Simulation}  \label{SIMUL}


The fluctuation in the 21-cm brightness temperature may depend on many
factors, such as the gas density, temperature, neutral fraction,
radial velocity gradient and $\rm Ly\,\alpha$ flux
\citep{2005ApJ...624L..65B}. In our work, we do not consider the redshift space 
distortion effect caused by the non-zero radial peculiar velocity gradient, and 
simply assume the
brightness temperature is proportional to the density of the neutral
gas.  
\beq
T_b\approx(27{\mK}){\left(\frac{1+z}{10}\right)}^{1/2}\frac{T_s-T_{\rm CMB}}{T_s}(1+\delta_{\rm HI})
\ , 
\eeq 
where $T_b$ is the brightness temperature increment respective
to CMB, $T_s$ is the spin temperature, which is about to be much bigger than
$T_{\rm CMB}$, and $\delta_{\rm HI}$ is the over-density of the neutral hydrogen.

Our work mainly focuses on the non-Gaussian aspect and 3-D properties of 
the reconstruction,
and these effects also exist in a pure dark matter distribution.
The neutral gas will trace the total mass distribution, which is
 dominated by the dark matter haloes. A simplification is to
use the dark matter as the source directly.  Even though this will
bring some bias at small scales, the approximation is valid at large
scales. The dark matter distributions are generated using the PMFAST
code \citep{2005NewA...10..393M}.


The high resolution PMFAST simulation was performed on a
$1456^3$ fine mesh with $3.9\times10^8$ particles. The production
platform was the IA-64  'lobster' cluster at CITA, which consists of 8
nodes. One of them  was  upgraded, so we have used the remaining 7
nodes. Each node contains four 733 MHz Itanium-1 processors and 64 GB
RAM. The simulation started at an initial redshift $z_i=100$ and ran
for 63 steps with comoving box-size $L=50h^{-1}\,\Mpc$.  The initial
condition was generated using the Zeldovich approximation, and the
matter transfer function was calculated using CMBFAST
\citep{1996ApJ...469..437S}.  The cosmological parameters were chosen
in accordance with the WMAP result \citep{2003ApJS..148..175S}:
$\Omega_{\rm m}=0.27, \Omega_{\Lambda}=0.73, \Omega_b=0.044,n=1.0,
\sigma_8=0.84$, and $h_0=0.71$. 20 independent boxes were generated.
We had 3-D data at $z=7$ at hand, and used them in our numerical tests for 
convenience.

\subsection{Convergence and shear map construction} \label{NUM_RC}

The dimensionless power spectrum, which is
the contribution to the variance of over-density per logarithmic
interval in spatial wave number, can be measured from the source data
in the periodic simulation box.

\begin{figure*}
\psfig{figure=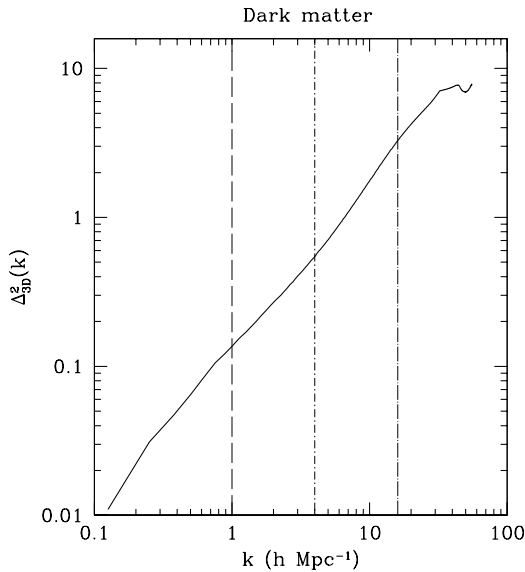,width=80mm,angle=0.}
\caption{ 
The dimensionless power spectra of the re-sampled dark matter
from the $1456^3$ N-body simulation in three dimensions are given.
The solid line is the power spectrum on the $512^3$ grids. The
re-sampled sources keep  
the non-linearity and the non-Gaussianity of the structures up to
$k\sim 30h\,\Mpc^{-1}$. 
Three different experimental noise cut offs are shown
with $k_{\rm c}=1h\,\Mpc^{-1}$,$4h\,\Mpc^{-1}$,
$16h\,\Mpc^{-1}$, which represent the linear, quasi-linear and non-linear scales.
}
\label{fig:ps3d}
\end{figure*}

To reduce the computation 
time, our numeric results on the reconstruction use a re-sampled distribution. 
We will generate 20 independent sources each on $512^3$ grids, to 
investigate the statistics. 
The total co-moving length along the line-of-sight of 20 simulation boxes is 
$1 h^{-1} \,\rm Gpc$, which is about the same size as the observable 21-cm region 
distributed between redshifts $10-20$.
The correlation  between the boxes can be ignored since the box-size is
much larger than the non-linear length scale, and the number of
neglected modes is small.
In Fig. \ref {fig:ps3d}, the solid line is the power spectrum of the
re-sampled sources.  To measure the dependence of non-Gaussianity on
scale, we compare the results with different scales of experimental
noise cut off.  

We simply assume the noise to be zero above a cut off and 
infinity below the cut off scale. This is a reasonable approximation
for a filled aperture experiment, which has good brightness
sensitivity, and an exponentially growing noise at small scales.  Three
cut off where chosen at $k_{\rm c}=1h\, \Mpc^{-1}$, $4h\, \Mpc^{-1}$,
$16h\,\Mpc^{-1}$, which represent the linear, quasi-linear and
non-linear scales.  Three different experimental noise levels are shown
as vertical lines in Fig. \ref {fig:ps3d}.

In principle, the convergence map is the variance (or covariance when the
filter ${\cal F}^{\rm \kappa}$ has negative value) of
the over-density field after a specified filtering process.  
Shear is the covariance of two maps, since the anisotropic filter can
not be factored in to a perfect square. We need to smooth the
maps to extract the lensing signal with  maximum S/N. The window
function used to smooth the lensed map, which is isotropic in the
transverse directions to the line-of-sight, can be calculated with
Eq. (\ref {eq:wk_pk}).  The gradient of the power spectrum
becomes negative at small scales; that comes from the limited
resolution of the N-body simulation and is unphysical. The
experimental noise will put a natural cut off at small scales.

As mentioned in Section $2.2.2$, the reconstruction of $\kappa$ will
depend on $2P+\Delta P$, in 2-D which is equivalent to the gradient of
2-D version of $\Delta^2_{\rm 2D}=k^2 P_{\rm 2D}(k)/2\pi$. In 3-D, it
is more complicated since $\Delta P(\k)$ is not isotropic.  The
optimal window functions have two parts $W_1$ and $W_2$, the choice of
which is not unique. One might expect a symmetric
decomposition to have the best S/N.  The optimal filter of
$\kappa$ is positive except at a few modes, and can be decomposed in to two equivalent parts (one part need to contain a minus sign for those negative value of the filter).  In contrast to $\kappa$, the shear construction needs to use the
covariance between two different windowed temperature fields, since
there is a $\sin$ or $\cos$ component in the window function.  The
window is a function of the transverse and parallel components of
$\k$.  

We can calculate the mean covariance of the two smoothed maps along the redshift axis for each pixel.  From Eq. (\ref{eq:kappax}) we can construct the
convergence map.  Shear maps are reconstructed in the same way, except
different optimal window functions are used.  
The anisotropic part $\cos{2\theta_{\kt}}$
can be decomposed into $\cos{\theta_{\kt}}- \sin{\theta_{\kt}}$ and
$\cos{\theta_{\kt}}+ \sin{\theta_{\kt}}$.  Both windows can generate a
field with even PDF so that the distribution is less
non-Gaussian. This is consistent with the numerical results as shown
in Fig. \ref{fig:PDF}.  Using these two maps, we construct
the $\gamma_1$ map with their covariance, as shown in the
Eq. (\ref{eq:shearx}). Similarly we can get the $\gamma_2$ map.

\section{Numerical results and discussion} \label{DISC}
\begin{figure*}
\psfig{figure=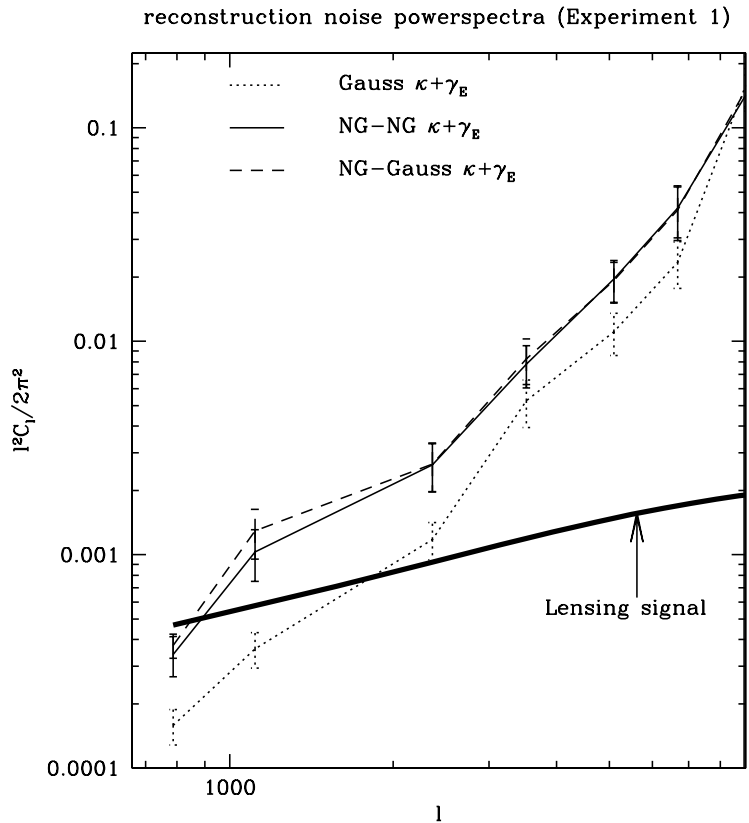,width=120mm,angle=0.}
\caption{The noise of lensing maps from different estimators using
 experimental noise 1, which cuts off at $k_c=1h\,\rm Mpc^{-1}$. 
We treat the $1h^{-1}\,\rm Gpc$ space of gas at $z=10-20$ as 20 independent sources each is a $50h^{-1}\,\rm Mpc$ box-size cube. 
Structures at these redshifts are similar to those at $z=7$ used by us, 
though less non-linear. We can expect to see similar non-Gaussianity effects in the reconstruction 
with the $1h^{-1}\, \rm Gpc$ space except that the non-Gaussianity of sources 
will be smaller.
The curves are truncated at $\sqrt{2} k_{\rm c}$, where the noise goes
to infinity.  The thick solid line is the expected lensing signal.
The dotted line is the lensing reconstruction noise
for a simulated Gaussian source with the same power spectrum.  The
dashed curve is the noise from the N-body simulation using the
Gaussian estimator, which increases modestly compared to the Gaussian
source.  It is identical for the optimal $\kappa,\gamma$ 
reconstruction as it is for the deflection angle.   The thin solid
line is noise when shear and convergence are re-weighted by their
non-Gaussian variances.
}
\label{fig:noise_Exp1}
\end{figure*}
\begin{figure*}
\psfig{figure=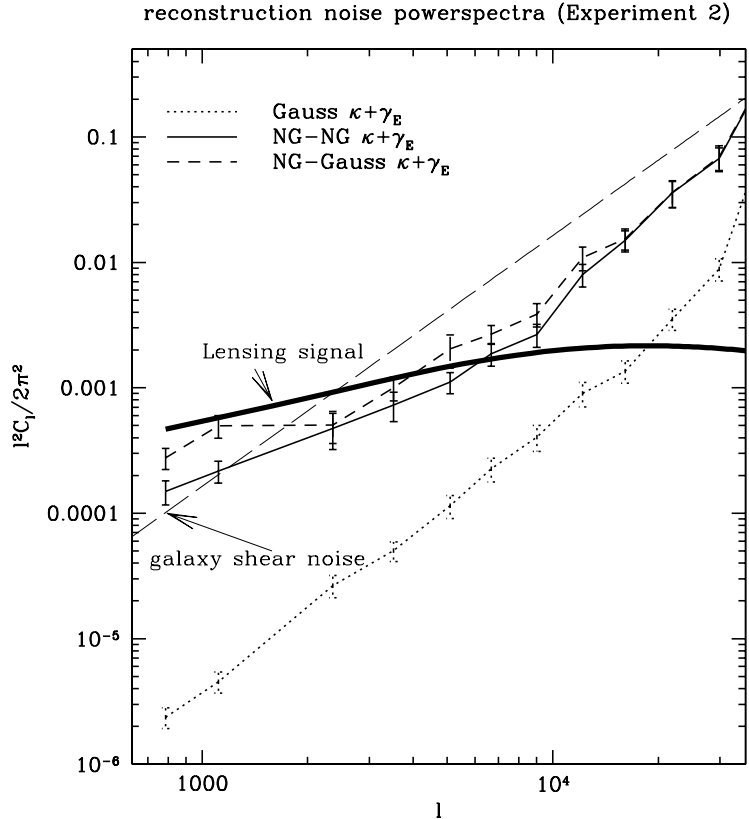,width=120mm,angle=0.}
\caption{
Same of Fig.~\protect\ref{fig:noise_Exp1}, but with cut off at the
quasi-linear scales $k_c=4 h\, \rm Mpc^{-1}$. 
The effect of non-Gaussianity of sources is more pronounced.
We can compare the S/N with a fiducial cosmic shear survey of sources
in the same $10<z<20$ redshift range, which reconstructs the lensing
from the shape of galaxies, with a surface density of $14\,\rm
arcmin^{-2}$.  To map the lensing to the same S/N with
redshift $z\sim 1$ sources requires a density of  $56\,\rm
arcmin^{-2}$ \citep{2001ApJ...554...67H} with rms ellipticity of 0.4.
We see that proposed optical lensing surveys are unlikely to
outperform 21-cm sources.
}
\label{fig:noise_Exp2}
\end{figure*}
\begin{figure*}
\psfig{figure=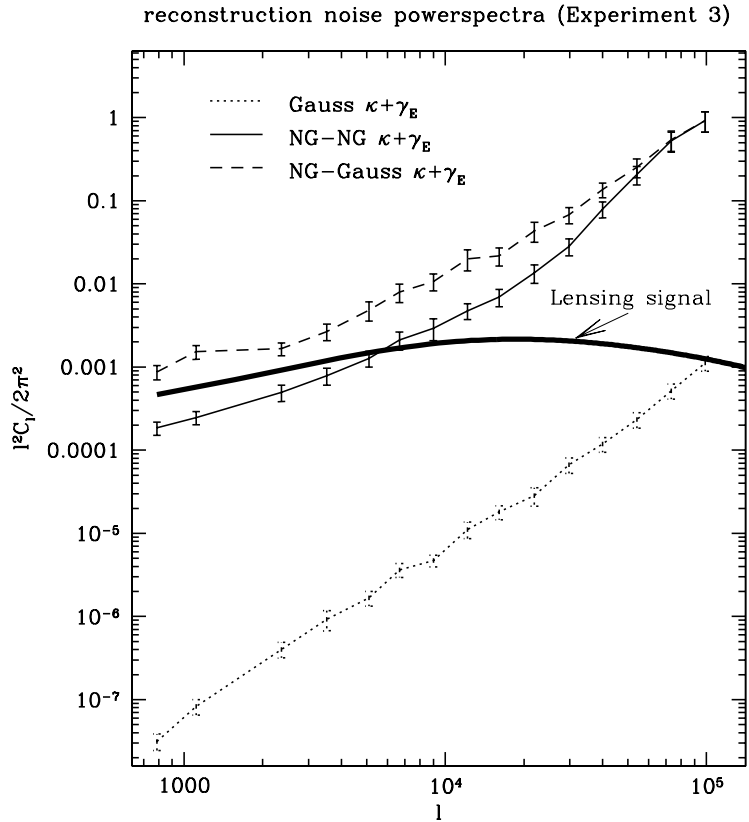,width=120mm,angle=0.}
\caption{Same of Fig.~\protect\ref{fig:noise_Exp1}, but with cut off at the
non-linear scales $k_c=16 h\, \rm Mpc^{-1}$.
At the highly non-linear scales, the non-Gaussian noise is about 3 to
4 magnitude higher than the Gaussian noise. 
The combined re-weighted estimator (NG-NG
$\kappa+\gamma_{\rm E}$) has noise about half an order of magnitude lower than the OQDE.
}
\label{fig:noise_Exp3}
\end{figure*}
\begin{figure*}
\psfig{figure=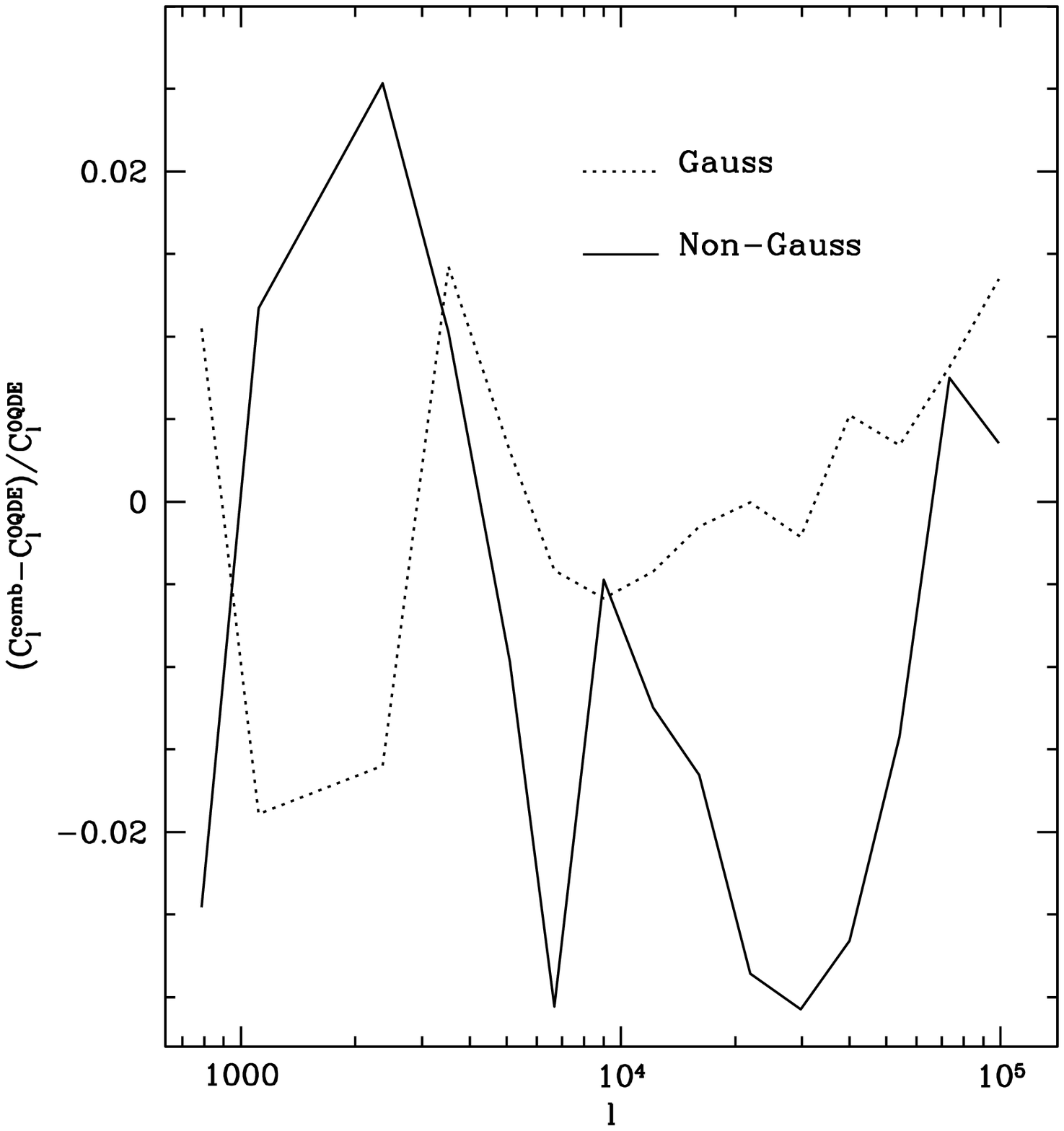,width=120mm,angle=0.}
\caption{The comparison of reconstruction noise from the combined
$(\kappa,\gamma)$ estimator and the OQDE.  While the
optimality is only proved at low $l$, we find them equally optimal
for Gaussian sources at all scales. 
The scatter is
consistent with numerical integration errors from the tabulated power
spectrum.
}
\label{fig:Comb_OQDE}
\end{figure*}

\citet{2004NewA....9..173C} claims that the variance will not vary
considerably and is not a ideal measurement of the lensing signal.
Even though the $\kappa$ field itself is only a few percent,
the integrated effect from the 3-D images will reduce the noise
ratio significantly to uncover the signal. \citet{2006ApJ...653..922Z}
solve the problem from an alternative approach by generalizing the
minimum variance quadratic estimator \citep{2002ApJ...574..566H} in
CMB lensing to 3-D.  

A related work was done in \citet{2006astro.ph.11862B}, where they
also construct quadratic estimators of shear and convergence in real
space, even though they did not include the correlation between the
2-D slices along the line-of-sight and they did not choose the
estimator with minimized noise.

\subsection{Non-Gaussianity} \label{NG}

The dark matter distribution is linear at large scales, and can be treated as
Gaussian. In the non-linear scales, when the amplitude of density
fluctuations is big, the structure becomes highly non-Gaussian.
Reference Gaussian sources with identical power spectrum to the dark matter 
are generated. 

We treat the $1h^{-1}\,\rm Gpc$ region at $z=10-20$ as 20 independent sources. 
Structures at these redshifts are similar to those at $z=7$ used by us, 
though less  non-linear. We can expect to see similar non-Gaussianity effects 
in the reconstruction 
with the $1h^{-1}\, \rm Gpc$ space except that the non-linear scale  
is smaller.
We compare the reconstruction noise with three different experimental
noise as well as the lensing signal in Fig. \ref{fig:noise_Exp1},
\ref{fig:noise_Exp2} and \ref{fig:noise_Exp3}. 
The thick solid line in the middle panel is the
lensing power spectrum, which is calculated with the Limber integral
of the 3-D power spectrum of dark matter using Eq. (\ref{eq:Limber}). We use the
publicly available code Halofit.f \citep{2003MNRAS.341.1311S} to
generate the nonlinear dark matter power spectrum. The code provides
both their fitting results, and the results using the
Peacock-Dodds formula (\citet{1996MNRAS.280L..19P}, PD96
hereafter). The 'stable clustering' assumption of PD96 breaks down at
low redshifts, but is reasonably good at high redshifts where the
power spectrum is more linear. The halofit code fits the power spectrum
at low redshift to Virgo and GIF CDM simulations, which used the
transfer function of \citet{1992MNRAS.258P...1E}. We use a
combination of the two: Halofit power spectra are used for redshifts
lower than $z=3.0$, and PD96 power spectra are used for higher
redshifts.

Since the reconstruction noise of $\kappa$ is isotropic, one can
always choose the direction of the lensing mode $\l$ to be parallel
with a coordinate axis. In this direction, $\gamma_1(\l)=\kappa(\l),
\gamma_2(\l)=0$, and $\gamma_{\rm E}=\gamma_1$, which simplifies the
numerical calculation.  The optimal combined estimator becomes the
sum of $\kappa$ and $\gamma_{\rm E}$ weighted by their noise.  The
weights could be the Gaussian $\kappa,\gamma_{\rm E}$ noise, or
non-Gaussian noise. We will show that the combined estimator with
Gaussian noise weights has the same noise as the OQDE for both Gaussian
and non-Gaussian sources.  Fig. \ref{fig:noise_Exp1},
\ref{fig:noise_Exp2}, and \ref{fig:noise_Exp3} are results using noise
cut offs from experiment 1, 2 and 3.  The curves are truncated at
$\sqrt{2}k_{\rm c}$.  The non-Gaussianity increased the noise of all
estimators. The first cut off falls in the linear regime, where the
non-Gaussianity only has a modest effect on the noise. The second
cut off is at the quasi-linear scale.  Here the non-Gaussianity
increases the noise of the OQDE by about 1 to 2 orders of magnitude. At
the highly non-linear scales, the non-Gaussian noise is about 3 to 4
magnitude higher than the Gaussian noise, and in fact higher than that 
for the more noisy experiment.  

Our estimators were derived in the limit that $\kappa$ and $\gamma$
are constant, and are optimal in that limit.  For spatially variable
lens, we solve for the required normalization factors.  In the OQDE, 
the windows do not depend on the scale of the lens, so one might
guess the same Ansatz to hold for the $(\kappa,\gamma)$ estimators.
We verify this numerically in Fig. \ref{fig:Comb_OQDE}.  The solid line
and dotted line is for Gaussian sources and non-Gaussian sources
respectively.  The differences are less than a few percent, and
consistent with integration errors from the tabulated power spectrum,
and most importantly, independent of scale, as we had expected.   We
do note, that for a finite size survey, the  $(\kappa,\gamma)$ recover
the constant mode, which is lost in the OQDE.  Three more
numbers are recovered.

The combined estimator with $\kappa$ and $\gamma_{\rm E}$ weighted by
non-Gaussian noise is more optimal than weighted by Gaussian noise,
therefore has lower noise than the OQDE.  In fact, the non-Gaussian noise
of $\gamma_{\rm E}$ is much smaller than $\kappa$.  To investigate the
origin of this change, we first investigate the cause of the increased noise
in non-Gaussian sources for $\kappa$.  This could be because either
the non-Gaussianity leads to a high kurtosis in $\kappa$, which boosts
the errors;  or the non-Gaussianity may lead to correlations between
modes, resulting in a smaller number of independent modes, and thus a
larger error.

\begin{figure*}
\psfig{figure=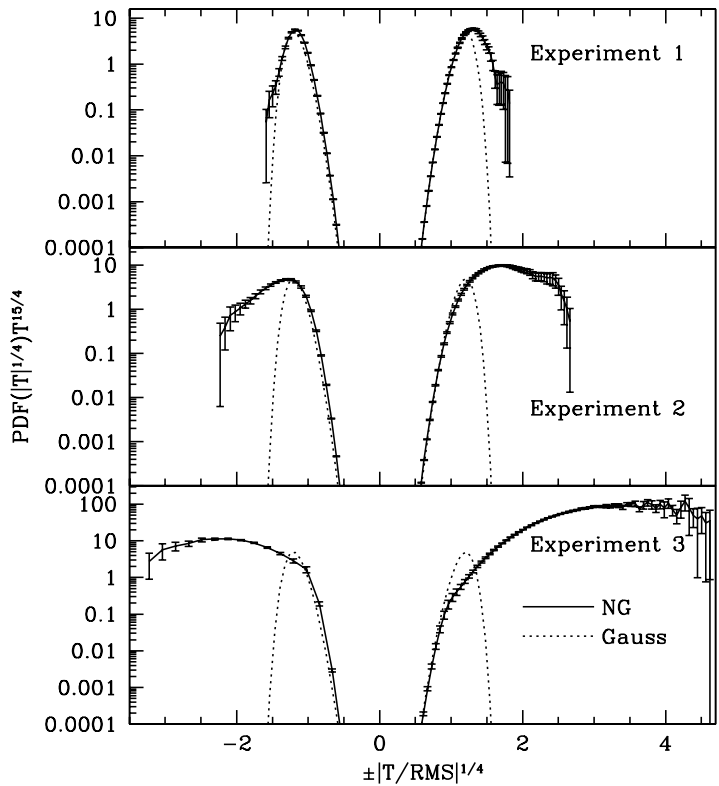,width=120mm,angle=0.}
\caption{
The PDF of maps smoothed
with $\kappa$ window are shown. The top, middle and bottom panel show
the results with experimental noise cut offs 1, 2 and 3. 
The solid line is the PDF for maps smoothed with the $\kappa$ window 
($T^{\kappa}$ in section \ref{EST_K}). 
To see the full dynamic range on the x-axis,
we plot the curve with $\pm |T|^{1/4}$ as x-axis, and 
${\rm PDF}(|T|^{1/4})|T|^{15/4}$ as the y-axis. The integral of the
x-axis weighted by the y-axis 
will give $\langle{T^4}\rangle$, which is basically a estimation of
the reconstruction noise.  The error bars are estimated from the 20
simulations. 
To compare with a Gaussian distribution, dotted lines are also plotted.
The contributions to the $\langle{T^4}\rangle$ in experiment
1 mainly come from small fluctuation regions. In experiment 2, the
large outliers play a more important role but one can still expect the
curve to converge. In experiment 3, most contributions come from rare
regions with high fluctuations.  Caution should be exercised in the
interpretation of the most non-linear scales, since a larger number of
source samples may result in a different error.  It is clear, however,
that the noise has increased dramatically.
}
\label{fig:PDF}
\end{figure*}

In Fig. \ref{fig:PDF}, the PDF of maps smoothed with the $\kappa$
window are shown. The top, middle and bottom panel show the results
with experimental noise cut offs 1, 2 and 3.  The solid line is the PDF
for maps smoothed with $\kappa$ window ($T^{\kappa}_1,T^{\kappa}_2$ in
section \ref{EST_K}).  Because the window functions are almost symmetric,
we plot only one PDF.  To see the full dynamic range on the x-axis,
we plot $\pm |T|^{1/4}$ as x-axis, and
${\rm PDF}(|T|^{1/4})|T|^{15/4}$ as the y-axis. The integral of the
x-axis weighted by the y-axis will give $\langle{T^4}\rangle$, which
is basically a estimation of the point-wise non-Gaussian reconstruction
noise. Here ${\rm PDF}(|T|^{1/4})$  is the PDF of $|T|^{1/4}$. To compare with a Gaussian distribution, dotted lines are also
plotted.  The contributions to the $\langle{T^4}\rangle$ in experiment
1 mainly come from small fluctuation regions. In experiment 2, the
large outliers play a more important role but one can still expect the
curve to converge. In experiment 3, most contributions come from rare
regions with high fluctuations.  Caution should be exercised in the
interpretation of the most non-linear scales, since a larger number of
source samples may result in a different error.  It is clear, however,
that the noise has increased dramatically.

The kurtosis of $\kappa$ is
$\langle{(T^{\kappa}_1)^4}\rangle/\langle{(T^{\kappa}_1)^2}\rangle^2-3$,
and an analogous quantity can be defined by $\langle{(T^{\gamma}_1
  T^{\gamma}_2)^2}\rangle/ (\langle{(T^{\gamma}_1)^2}\rangle
\langle{(T^{\gamma}_2)^2}\rangle) -1 $ for shear.
$T^{\kappa}_1\approx T^{\kappa}_2$, and $T^{\gamma}_1$ is uncorrelated with
$T^{\gamma}_2$.  The noise of $\kappa$ and $\gamma$ is determined by
both kurtosis and number of independent cells.
For experimental noise 1, the kurtosis of $T^{\kappa}$ and $T^{\gamma}$ are
$1.2$ and $0.29$ respectively. 
The effectively independent cube cells for $\kappa$ and $\gamma$ have side length 
$4.8 h^{-1}\,\rm Mpc$ 
and $4.6h^{-1}\,\rm Mpc$ respectively. The corresponding Gaussian sources with
the same cut off have effective cell size $3.0h^{-1}\,\rm Mpc$ and $3.5h^{-1}\,\rm
Mpc$.  
For experimental noise 2, the kurtosis of $T^{\kappa}$ and $T^{\gamma}$ are
$18$ and $5.7$ respectively. 
The effective cell size for $\kappa$ and $\gamma$ are $1.8h^{-1}\,\rm Mpc$ 
and $1.5h^{-1}\,\rm Mpc$ respectively.
The corresponding Gaussian sources with the same cut off have effective cell size $1.0h^{-1}\,\rm Mpc$ and $1.1h^{-1}\,\rm
Mpc$.
For experimental noise 3, the kurtosis for $T^{\kappa}$ and $T^{\gamma}$ are
$1.6\times10^3$ and $3.5\times10^2$ respectively. 
The effective cell size for $\kappa$ and $\gamma$ are $540h^{-1} \,\rm Kpc$ 
and $310h^{-1}\,\rm Kpc$ respectively.
The corresponding Gaussian sources with
the same cut off have effective cell size $240h^{-1}\,\rm Kpc$ and $290h^{-1}\,\rm
Kpc$.
We conclude that the shear measurements have lower non-Gaussian noise
both because of a smaller point-wise kurtosis, and less correlation
between modes.

We will see later that experiment 2 has the largest S/N, which is
larger than unity for $l\lesssim 6000$.  We can compare the S/N with
cosmic shear surveys, which reconstruct lensing from the shape of
galaxies. The noise can be estimated by $\langle{\gamma^2}\rangle
/n_{\rm eff}$ \citep{2006ApJ...647..116H,2001ApJ...554...67H}, where
we use $\langle{\gamma^2}\rangle^{1/2} \approx 0.4 $ as the rms intrinsic
ellipticity, and $n_{\rm eff}$ is the effective number density of
galaxies.  We plot the shear noise from a survey of sources in the
same redshift range $10<z<20$ in Fig.~\ref{fig:noise_Exp2}, with a
surface density of $14\,\rm arcmin^{-2}$.  For more realistic source
redshifts $z\sim 1$ in proposed optical surveys
\citep{2001ApJ...554...67H}, this corresponds to a surface density of
$56\,\rm arcmin^{-2}$.  In the CFHTLS wide survey the source galaxies
are distributed at redshifts lower than 3, and their effective number
density is $\sim 12$ galaxies $\rm arcmin^{-2}$
\citep{2006ApJ...647..116H}.  This noise is larger still.  Even though
non-Gaussian 21-cm lensing saturates lensing reconstruction, it still measures more
modes than current proposed optical surveys.

\begin{figure*}
\psfig{figure=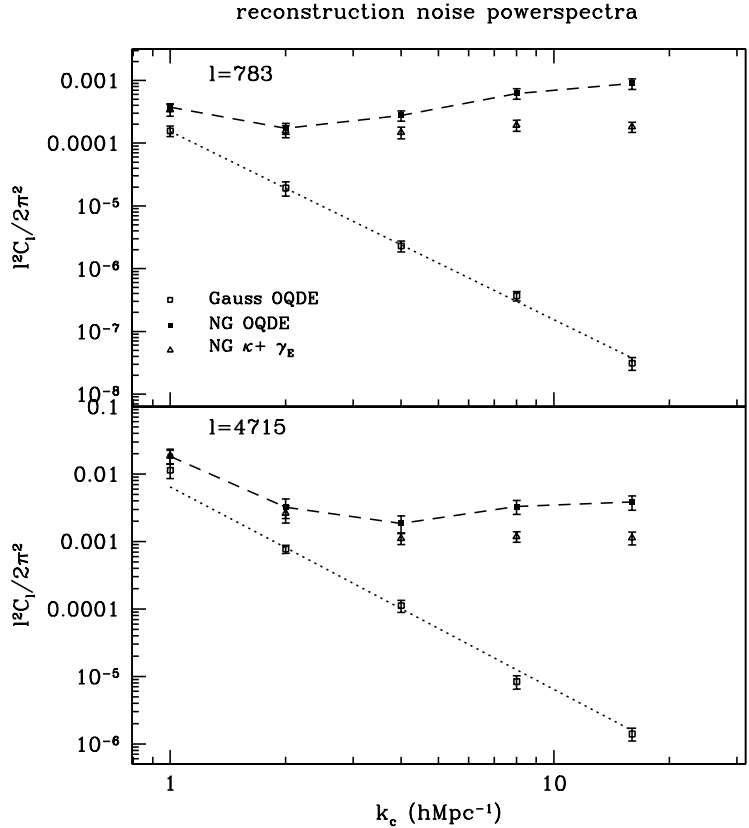,width=120mm,angle=0.}
\caption{ The reconstruction noise versus the cut off in the
experimental noise. The top  
panel is for $l_1=2\pi/L=783$, and the bottom panel is for
$l_2=6 l_1=4715$. The noise  
of Gaussian sources decreases as $k_{\rm c}$ increases, because of the
increasing number of independent modes.
The dotted lines are a least squares fitting power law $N_0 k_{\rm c}^{-3}$
to the Gaussian noises, and $N_0=3.1\times10^{-2},1.3\times10^{-1}$
for top and bottom panels respectively. 
The dashed lines connect the non-Gaussian noise of the OQDE.
The triangles are the reconstruction noise for the combined estimator, which
is equal to the OQDE at larger scale $k_{\rm c}$ and about half an
order of  magnitude lower at large $k_{\rm c}$. 
The noise of the non-Gaussian sources changes slowly and 
saturates or even increases at small scales. 
} 
\label{fig:noise_scale}
\end{figure*}

In Fig. \ref{fig:noise_scale}, we show the reconstruction noise at two
different $l$ versus various experimental noise cut off $k_{\rm c}$.  The
top panel is for the fundamental mode in the box, $l_1=2\pi/L=783$, and
the bottom panel is for $l_2=6l_1=4715$.  As shown in the
plot, it is clear that the noise of a Gaussian source decreases as
$k_{\rm c}$ increases, because of the increasing number of
independent modes.  The dotted lines are a least squares fitting power law
$N_0 k_{\rm c}^{-3}$ to the Gaussian noises, and
$N_0=3.1\times10^{-2},1.3\times10^{-1}$ for top and bottom panels
respectively.  This comes from counting the number of available source
modes.  The dashed lines connect the non-Gaussian noises of the
OQDE. The triangles are the reconstruction noise for the combination
estimator, which is equal to the OQDE at larger scale $k_{\rm c}$ and
about half an order of magnitude lower at smaller scales of $k_{\rm
  c}$.  From this plot, we can see that experiment with lower noise
does not necessarily decrease the reconstruction noise of the OQDE for
non-Gaussian sources. And the experimental noise has a limit around the
quasi-linear scale where the OQDE achieves its best S/N.  The S/N achieves
its maximum around $k^{\rm NG}_{\rm c}\approx 4h\,\Mpc^{-1}$. This
cut off with maximum S/N varies only slowly with $l$.  

If one wants to estimate the effective number of available lensing
modes, we can derive an effective cut off of a Gaussian field which gives
the same S/N as the optimal non-Gaussian sources estimator.  This is
$k^{\rm G}_{\rm c}\approx2h\,\Mpc^{-1}$, where the power spectrum of
source is $\Delta^2\approx 0.2$. 
The size of the effectively independent cells is $2.0 h^{-1}\,\rm Mpc$.
 A simple Gaussian noise estimate
counts all modes up to  $\Delta^2(k)< 0.2$, which is perhaps
surprisingly low.

For our noise estimates, we stacked simulations all at redshift
$z=7$.  While the angular diameter distance does not change much
to $z \sim 20$, the structure does evolve.  We do not have
access to the higher redshift outputs to test this effect, but one
would expect a smaller non-linear scale to result in a smaller
reconstruction noise.

\subsection{Future directions}

A possible way to find the optimal window functions for non-Gaussian
sources is to divide the window into $N$ frequency bins
$W_1(\k_1,\k_2,...,\k_N)$, and apply a numerical variation to
those bins.  The noise can be measured numerically by applying the
estimator to the simulated sources.  The process of searching for a
optimal filter is equivalent to look for a minimum of reconstruction
noise in $N$ dimensional space $\k_1,\k_2,...,\k_N$. One could use
a Newton--Raphson  method to do this.  In this paper we
only considered the class of windows which are identical to the
optimal Gaussian estimators with a hard cut off, as well as two
weightings for shear and convergence.

One can also try to Gaussianize the sources by modifying the PDF of all
the sources to be Gaussian. The physical explanation and details of
Gaussianization can be found in \citet{1992MNRAS.254..315W}.  The
basic idea is that every pixel should preserve its rank in the whole
field during the Gaussianization process. During structure formation,
the non-linear evolution at small scales should not destroy most of
the information on the peaks and dips of the linear field. However,
this Gaussianization process will change the power spectrum of sources,
and the reconstructed lensing field will be biased.  This is not a
linear process, and the variation of power spectrum does not have
analytical solution, and can only be measured numerically with
simulated sources.

Recently it has been proposed that one could economically achieve
brightness mapping of 21-cm emission at lower redshifts \citep{2007arXiv0709.3672C}, 
potentially even with existing telescopes.  If
individual galaxies are not resolved, one can again ask the question
of how one could reconstruct a lensing signal.  This is very similar
to the problem studied in this paper.

\section{conclusion} \label{CONC}

In this paper, we developed the maximum likelihood estimator for the
large-scale structure from the 21-cm emission of the neutral gas
before the epoch of re-ionization. The convergence and shears can be
constructed independently.  To test the effects of non-Gaussianity, we
applied our estimators to simulated data.  The sources were generated by 
N-body simulations, because gas is expected to trace the total mass
distribution.  To investigate the influence of non-Gaussianity, we
also use Gaussian sources which have the same power spectrum as the
simulated sources.  We applied our estimator and the OQDE on both the
Gaussian and non-Gaussian sources.  Though our estimators are derived
in the simplified case of a constant convergence, the noise of our
combined estimator of convergence and shear are the same as the OQDE
for Gaussian sources.  For a finite
survey area, three extra constant modes can be recovered.  

The non-Gaussian nature of the source can increase the error bar by
orders of magnitude, depending on the experimental cut off scale. Shear
construction is affected less by non-Gaussianity than the convergence
field, and the combined estimator with non-Gaussian noise weights is a
better choice than reconstructing with the OQDE.  S/N can not be
boosted infinitely by reducing the experimental noise, and achieves its
maximum for a cut off around $k^{\rm NG}_{\rm c}\approx 4h\,\Mpc^{-1}$.
Below that scale the S/N start to saturate or even decrease. The
maximum S/N for non-Gaussian sources is equal to Gaussian
sources with $k^{\rm G}_{\rm c}\approx2h\,\Mpc^{-1}$, where the power
spectrum of source is $\Delta^2\approx0.2$ and the side length of the effectively 
independent cells is $ 2.0 h^{-1}\,\rm Mpc$. The maximum S/N is
greater than unity for $l\lesssim 6000$, which makes 21-cm lensing very
competitive compared to optical approaches.

{\it Acknowledgments}
We thank  Oliver Zahn, Chris Hirata, Brice M\'{e}nard and Mike Kesden for helpful 
discussions. T.T. Lu thanks  Pengjie Zhang, Zhiqi Huang, Hy Trac, and Hugh Merz for help 
in the early stage of the work. 

\bibliography{lubib}
\bibliographystyle{mn2e}

\appendix
\section[]{ normalization and noise of the estimator}

In the end of section  \ref{DISC}, the numerical results of the noise of 
the estimators are shown. Here we will develop the analytical expression for   
\beq
E_{\rm\kappa}(\kt) = b_{\rm \kappa}(\kt) [\kappa (\kt)+n(\kt)]  \,.
\eeq
 For shear, a similar relationship holds even though 
$b$ and $n$ are not isotropic. 

When $\kappa$ is spatially variable, 
\beq
\tilde T_b(\x)
= T_b(\xt-\bmath D(\xt),\xp) 
= T_b(\xt,\xp)-\gradt T_b(\xt,\xp)\cdot \bmath D(\xt) \,,
\eeq
where $\bmath D(\xt)=\bmath d(\xt)\chi(z_s)$, and $\bmath d(\xt)$ is the deflection angle. 
Therefore $\kappa=\gradt \cdot \bmath D$.

Fourier transforming Eq. (\ref{eq:kappax}),

\beq 
E_{\rm \kappa}(\kt)
= \int \dtx {E_{\rm \kappa}(\xt)} e^{-{\rm i} \kt \cdot \xt } 
= {1\over L} \int \dvx {\tilde T^{\rm \kappa}_{\rm w_1}}(\x) 
    {\tilde T^{\rm \kappa}_{\rm w_2}} (\x) e^{-{\rm i} \kt \cdot \xt} 
- (2\pi)^2 \delta^{\rm 2D}(\kt)V_{\rm \kappa}\ .
\label{eq:kappak}
\eeq
$\tilde T=\tilde T_b+n$, and noise is uncorrelated with the signal. The product in real 
space can be represented as a convolution in Fourier space 
\beq
 \int\dvx e^{-{\rm i} \kt \cdot \xt } {\tilde T^{\rm \kappa}_{\rm w_1}}(\x) 
    {\tilde T^{\rm \kappa}_{\rm w_2}} (\x)   
= \int{\frac{{\rm d}^3k'}{(2\pi)^3}}  
{\tilde T^{\rm \kappa}_{\rm w_1}}(\kt',\kp')  {\tilde T^{\rm \kappa}_{\rm w_2}}(\kt-\kt',-\kp') \,.
\eeq

\beq
\tilde T_b(\k)
= \int\dvx  e^{-{\rm i} \k \cdot \x}  T_b(\xt-\bmath D(\xt),\xp)  
= T_b(\k) -\int \dvx e^{-{\rm i} \k \cdot \x} \gradt T_b(\xt,\xp)\cdot \bmath D(\xt) \,,
\eeq
and the lensing introduced term can be further simplified as 
\beqa
 \int \dvx e^{-{\rm i} \k \cdot \x} \gradt T_b(\xt,\xp)\cdot \bmath D(\xt) 
&=& \int \dvx e^{-{\rm i} \k \cdot \x} T_b(\xt,\xp) ({\rm i}\kt-\gradt)\cdot \bmath D(\xt) \nonumber\\ 
&=& \int {\frac{{\rm d}^2k_{\perp}'}{(2\pi)^2}} T_b(\kt-\kt',\kp)({\rm i}\kt-\gradt)\cdot \bmath D(\kt') \,. 
\eeqa 
The quadratic term in Eq. (\ref{eq:kappak}) can be written as
\beqa
 \int\dvx e^{-{\rm i} \kt \cdot \xt}  {\tilde T^{\rm \kappa}_{\rm w_1}}(\xt,\xp) 
    {\tilde T^{\rm \kappa}_{\rm w_2}} (\xt,\xp)   
&=& \int {\frac{{\rm d}^3k'}{(2\pi)^3}}
 W^{\rm \kappa}_1(\kt',\kp')W^{\rm \kappa}_2(\kt-\kt',-\kp') 
 [T_b(\kt',\kp') T_b(\kt-\kt',-\kp')  \nonumber \\
& &  - T_b(\kt-\kt',-\kp') \int {\frac{{\rm d}^2k_{\perp}''}{(2\pi)^2}} T_b(\kt'-\kt'',\kp') 
 ({\rm i}\kt'-\gradt)\cdot \bmath D(\kt'') \nonumber\\ 
& &  - T_b(\kt',\kp') \int{\frac{{\rm d}^2k_{\perp}'''}{(2\pi)^2}} T_b(\kt-\kt'-\kt''',-\kp') 
 \left({\rm i}(\kt-\kt')-\gradt\right)\cdot \bmath D(\kt''')] \nonumber\\
& & + {\rm Noise} \,.
\label{eq:kappak_long}
\eeqa

Using the relationship that
\beq
 \langle{T_b(\kt',\kp') T_b(\kt-\kt',-\kp')}\rangle =(2\pi)^3\delta^{\rm 3D}(\kt,0) P_{\rm 3D}(\kt',\kp') \ , 
\eeq
we found that the expectation value of the first terms and the noise term in 
Eq. (\ref{eq:kappak_long}) can cancel the last term in Eq. (\ref{eq:kappak}). 
Note $\delta(0)=\lim_{\Delta k \rightarrow 0} (\Delta k)^{-1}\sim ({L/ 2\pi})$, 
and $W^{\rm \kappa}_2(\kt-\kt',-\kp')\sim W^{\rm \kappa}_2(\kt',\kp')$ since $\delta^{\rm 2D}(\kt)$ is nonzero only
 when $\kt=0$. Similarly, the last two terms can be simplified.
The normalization factor
\beq
b_{\rm \kappa}(\kt) 
= {2\over k_{\perp}^2}  \int{\frac{{\rm d}^3k'}{(2\pi)^3}} W^{\rm \kappa}_1(\kt',\kp')W^{\rm \kappa}_2(\kt-\kt',-\kp') 
  [ (\kt-\kt')\cdot\kt P_{\rm 3D}(\kt-\kt',-\kp')+\kt'\cdot\kt P_{\rm 3D}(\kt',\kp') ] \,.
\eeq
Similarly, replacing $W^{\rm \kappa}_1,W^{\rm \kappa}_2$ by $W^{\rm \gamma_1}_1,W^{\rm \gamma_1}_2$
($W^{\rm \gamma_2}_1,W^{\rm \gamma_2}_2$), and $k_{\perp}^2$ by $k_{\perp}^2\cos{2\theta_{\kt}}$
($k_{\perp}^2\sin{2\theta_{\kt}}$), we find the normalization factor for $\gamma_1$
($\gamma_2$). 

The noise of the estimator 
can be calculated in the absence of lensing:
 $ \langle{|\hat \kappa(\kt)|^2}\rangle= \langle{\hat \kappa(\kt)\hat \kappa^{\star}(\kt) }\rangle$.
Since $\langle{|\hat \kappa(\kt)|^2}\rangle= (2\pi)^2 \delta^{\rm 2D}(0) N_{\kappa}(\kt)$
and $\delta^{\rm 2D}(0)=\lim_{\Delta k \rightarrow 0} (\Delta k)^{-2}\sim ({L/ 2\pi})^2$, 
Wick's theorem gives
\beqa
N_{\rm \kappa}(\kt) 
&=& {1\over b(\kt)^2 L} \int {\frac{{\rm d}^2k_{\perp}'}{(2\pi)^2}} \int {\frac{{\rm d}\kp'}{(2\pi)}}
 \{P_{\rm 3D}(\kt-\kt',-\kp')P_{\rm 3D}(\kt',\kp') 
   [W^{\rm \kappa}_1(\kt-\kt',-\kp')W^{\rm \kappa}_2(\kt',\kp')]^2 \nonumber \\
&+& P_{\rm 3D}(\kt-\kt',-\kp')P_{\rm 3D}(\kt',\kp'){\cal F}^{\rm \kappa}(\kt-\kt',-\kp') 
 {\cal F}^{\rm \kappa}(\kt',\kp') \} \,.
\eeqa
The first term is the convolution of $P_{\rm 3D}(\k)W^{\rm \kappa}_1(\k)^2$ and $P_{\rm 3D}(\k)W^{\rm \kappa}_2(\k)^2$, 
and the second term is the convolution of $P_{\rm 3D}(\k){\cal F}^{\rm \kappa}(\k)$ with itself. 
The dimensionless quantity $k_{\perp}^2 N_{\rm \kappa}(\kt)/(2\pi)$ is equivalent to $l^2C_l/(2\pi)$ in other literature.

\bsp
\label{lastpage}

\end{document}